\documentclass[pre,superscriptaddress,twocolumn,final,showpacs,showkeys,nobibnotes,titlepage,twoside,10pt]{revtex4-1}
\usepackage{amsmath}
\usepackage{graphicx}
\usepackage{amsfonts}
\usepackage{amssymb}
\usepackage{subfigure}
\usepackage{float}
\usepackage{color}

\begin{document}

\author{D. V. Denisov}
\affiliation{Van der Waals-Zeeman Institute, University of Amsterdam, The Netherlands}
\author{M. T. Dang}
\affiliation{Van der Waals-Zeeman Institute, University of Amsterdam, The Netherlands}
\author{B. Struth}
\affiliation{Deutsches Elektronen-Synchrotron, Hamburg, Germany}
\author{G. H. Wegdam}
\affiliation{Van der Waals-Zeeman Institute, University of Amsterdam, The Netherlands}
\author{P. Schall}
\affiliation{Van der Waals-Zeeman Institute, University of Amsterdam, The Netherlands}

\begin{abstract}
Yielding is central to the relaxation, flow and fracture of a wide range of soft and molecular glasses, but its microscopic origin remains unclear. Here, we elucidate the yielding of a colloidal glass by using x-ray scattering to monitor the structure factor during the yielding process. We apply a recently introduced combination of small-angle x-ray scattering and rheology to the oscillatory shear, and follow the structure factor during the increasing strain amplitude. Surprisingly, we observe a sharp transition in the orientational ordering of the nearest-neighbor structure upon yielding, in contrast to the smooth variation of the viscoelastic moduli. This transition is accompanied by a sudden change of intensity fluctuations towards Gaussian distributions. We thus identify yielding as a new, dynamically induced transition of the glass in response to the applied shear.
\end{abstract}

\title{Particle response during the yielding transition of colloidal glasses}
\maketitle


The yielding of glasses is important for a wide range of materials including metallic glasses, polymer- and soft glasses. Yielding demarcates the property of any solid to flow and deform irreversibly under applied deformation~\cite{Barnes1999}. At small applied stress and strain, the material deforms mostly elastically; at larger strain, the material starts to flow irreversibly, resulting in permanent deformation. Glasses are structurally frozen liquids with relaxation times exceeding the experimental time scale by many orders of magnitude and hence exhibiting solid-like properties~\cite{Ediger1996}. Yielding is central to many properties of glasses including time-dependent elasticity, relaxation, flow and fracture; insight into yielding should provide a deeper understanding of the glassy state~\cite{ColloidalGlasses}, but remains challenging. The yielding of glasses addresses an important fundamental question: how does the dynamically arrested state respond to the application of stress? While at the glass transition, microscopic observables change rather smoothly, yet rapidly~\cite{Pusey1987,Nagel1998} as a function of density or temperature, an important question to ask is whether a similarly smooth transition exists upon application of stress. The response to applied stress, however, remains poorly understood, partially because structural imaging of glasses during their yielding remains difficult~\cite{Petekidis2012}.

Colloidal glasses provide good models for a wide range of soft and simple molecular glasses; they exhibit dynamic arrest due to crowding at volume fractions larger than $\phi_g \sim 0.58$, the colloidal glass transition~\cite{Pusey1986,vanMegen}. Microscopically, the particles are trapped within cages formed by their nearest neighbors allowing only for very slow structural rearrangements. These systems exhibit glass-like properties such as non-ergodicity and aging \cite{Bouchaud1992}, and upon application of small but sufficiently large stress, they yield and flow~\cite{YieldingCollGlass}. The yielding of colloidal glasses has been widely investigated by oscillatory rheology, in which the sample is probed with a time-dependent, oscillatory strain. Yielding is usually associated with the intersection of the strain-dependent storage and loss moduli, but the exact definition of the yield point remains a matter of debate~\cite{Bonn2006,BonnDenn}. Constitutive relations have been used to model yielding based on structural parameters~\cite{Langer2011}, and recent advanced oscillatory rheology~\cite{Rogers2011} and combined rheology and simulation~\cite{Petekidis2012} have provided some rheological insight into the yielding process. Furthermore, mode-coupling theory and simulations~\cite{Brader2010} have studied the nonlinear stress response in oscillatory shear, and shear heterogeneities in steady state shear and creep~\cite{shearbanding_sim}. However, the direct observation and investigation of the structure during oscillatory yielding, which would give important insight into the nature of the yielding transition, remained elusive.

In this paper, we elucidate the microscopic yielding of glasses by direct measurement of the structure factor during the yielding process. We apply a recently introduced combination of x-ray scattering and rheology to the oscillatory shear of a colloidal glass to monitor the structure factor during the increasing applied strain amplitude. This allows us to obtain new insight into the nature of the yielding transition: We identify a sudden symmetry change in the orientational ordering, reflecting a surprisingly abrupt transition from a solid to a liquid-like state of the glass. Using a structural order parameter we demonstrate the sharpness of the induced transition as a function of the applied mechanical field. This sharp, dynamically induced transition under applied strain appears analogous to first-order equilibrium transitions.

\begin{figure}
\centering
\subfigure
{\includegraphics[width=0.6\columnwidth]{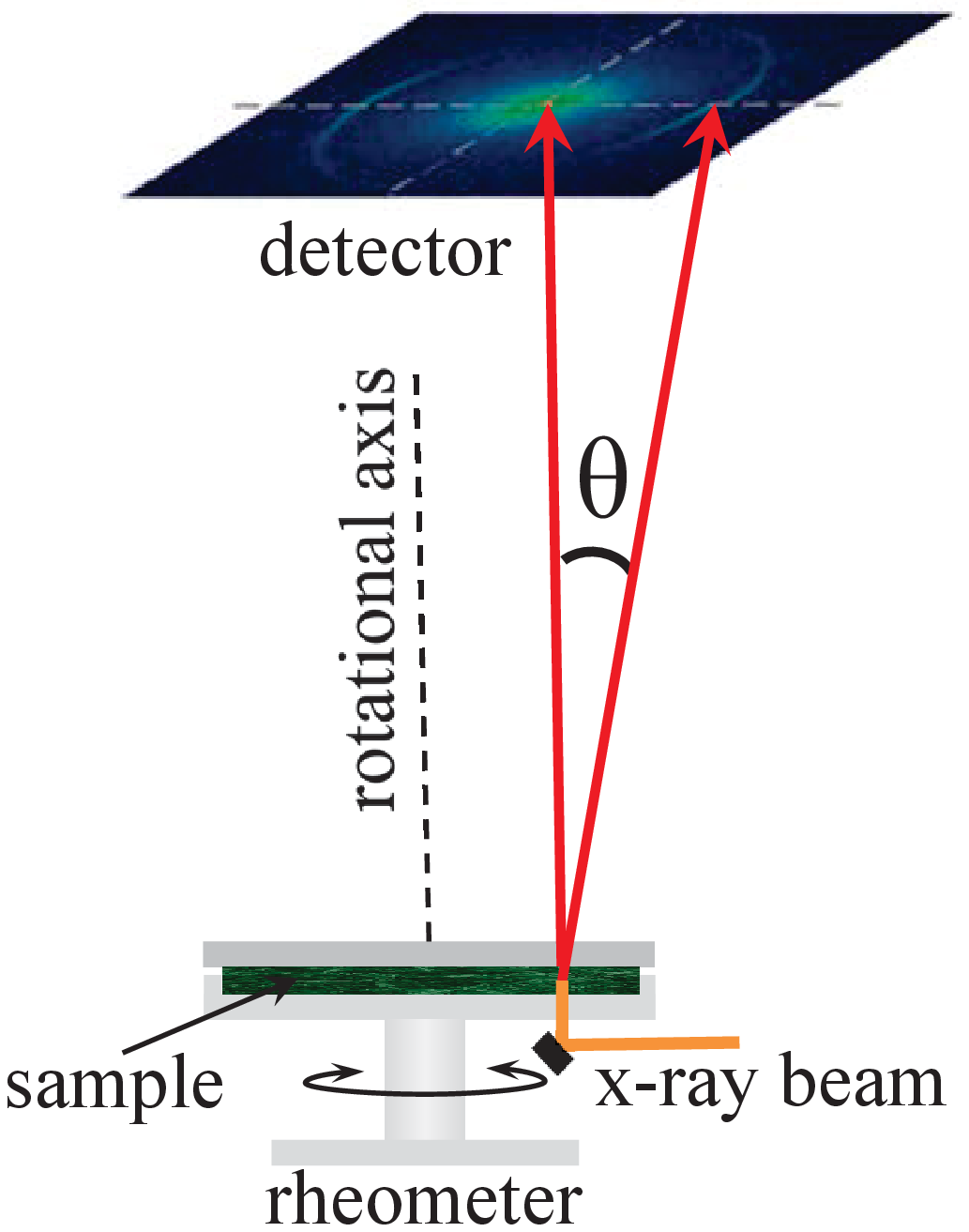}
\begin{picture}(0,0)(0,0)
\put(-155,0){(a)}
\end{picture}
\label{fig:Setup}}
\subfigure
{\includegraphics[width=0.85\columnwidth]{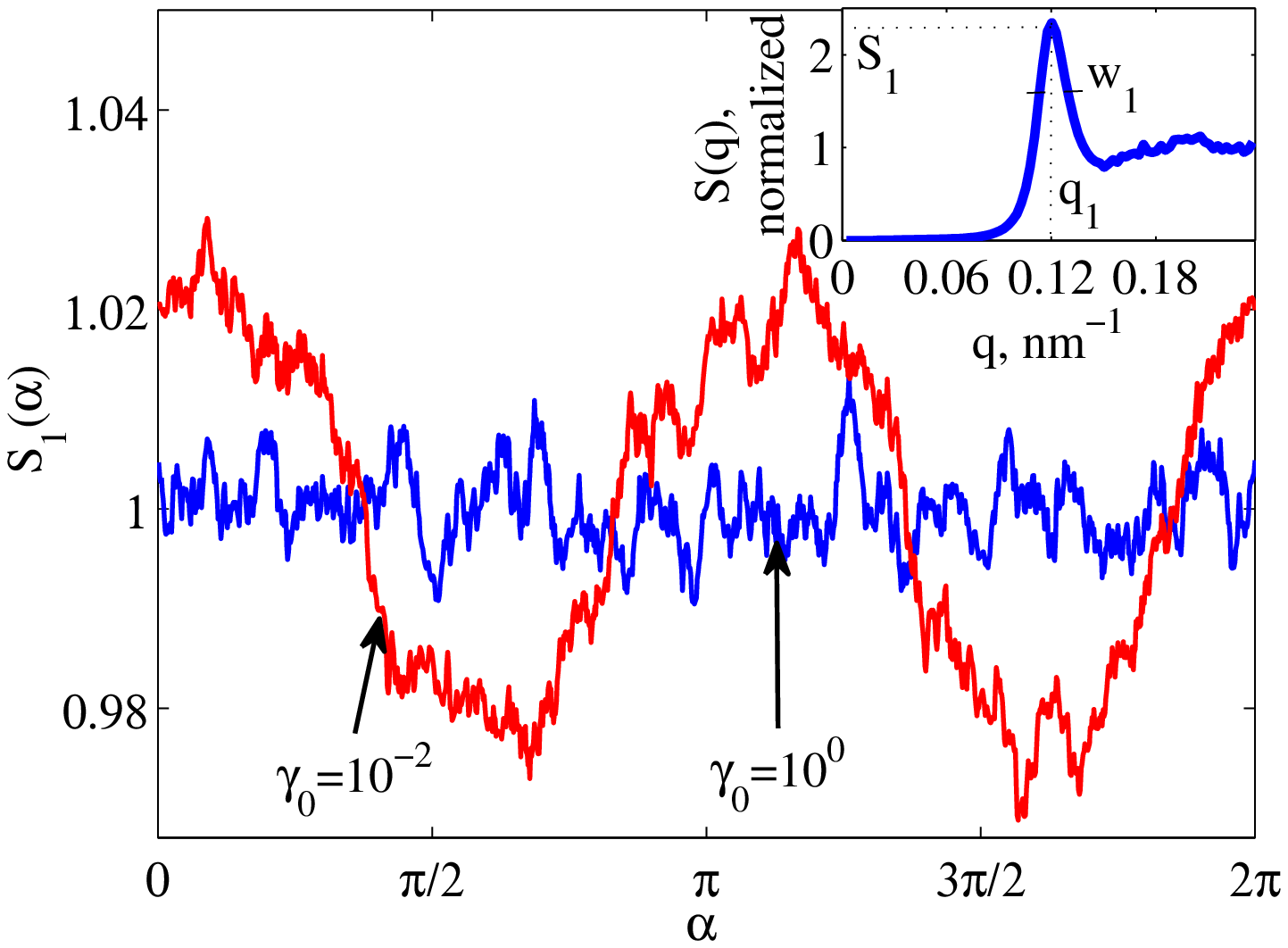}
\begin{picture}(0,0)(0,0)
\put(-200,0){(b)}
\end{picture}
\label{fig:Sq}}
\caption{(Color online)
\subref{fig:Setup} Schematic of the experimental setup showing the x-ray beam and detector with respect to the rheometer and the layer of sheared suspension. The rheometer is stress controlled and we use plate-plate geometry. The x-ray beam passes through the suspension at 0.78 times the disc radius; the beam diameter is smaller than 0.1 mm, much smaller than the disc radius of 18mm.
\subref{fig:Sq} Angle dependence of the first peak of the structure factor
$S(q_1)$ for small ($\gamma_0=10^{-2}$, red curve) and large strain amplitudes ($\gamma_0=10^{0}$, blue curve). Here, $\alpha$ is the angle with respect to shear direction. To calculate each value of $S(\alpha)$ we average in angular wedges of $\pi/30$ and radially over $\Delta q \sim2w_1$.
Inset: angle-averaged structure factor as a function of wave vector magnitude.}
\end{figure}

The experiments were carried out at the beam line P10 of the synchrotron PETRA III at DESY.  Simultaneous rheology and structure factor measurements were achieved by placing an adapted rheometer (Mars II, Thermo Fisher) directly into the beam path of the synchrotron~\cite{BerndLangmuir,Denisov2013}. The well-collimated synchrotron beam with wavelength $\lambda = 0.154$ nm is deflected into the vertical direction to pass through the layer of suspension perpendicular to the rheometer plates (Fig.~\ref{fig:Setup}).
The suspension consists of silica particles with a diameter of $50$ nm and a polydispersity of $10\%$, suspended in water. A small amount ($1 mM$) of $NaCl$ is added to screen the particle charges, resulting in a Debye screening length of $2.7$ nm, and an effective particle diameter of $2r_0 = 55.4$ nm.
Dense samples were prepared by diluting samples centrifuged to a sediment. Measurements of the relaxation time of these samples yielded $\tau \sim 10^6 t_B$~\cite{Denisov2013}, where $t_B$ is the Brownian time, indicating that the suspension is close to the colloidal glass transition~\cite{vanMegen1998}. Estimation of the effective volume fraction from dilution of the centrifuged sediment yields a value of $\phi \sim 58\%$, consistent with this interpretation~\cite{vanMegen1998}.
After loading, the samples are sealed with a small amount of low-viscosity oil to prevent evaporation and maintain sample stability over more than 4 hours. Samples were initialized by preshearing at a rate of $\dot{\gamma}=0.1$ s$^{-1}$ for 120 seconds followed by a rest time of 600 seconds; this procedure guaranteed reproducible results. We then apply oscillatory strain with frequency $f=1$ Hz and strain amplitude $\gamma_0$ increasing from $\gamma_{0min} = 10^{-4}$ to $\gamma_{0max} = 1$ (100 points on a logarithmic scale, three oscillations are averaged per cycle, leading to total duration of the experiment of around 5 minutes). We simultaneously monitor the scattered intensity using a Pilatus detector at a distance of $D = 280$ cm operating at a frame rate of 10 Hz. The detector with pixel size $172 \times 172$ $\mu$m$^2$ covers scattering angles $\theta$ in the range $0.03-0.5^{\circ}$, allowing access to wave vectors $q = 4\pi/\lambda sin(\theta/2)$ in the range $qr_0 = 0.5$ to $5$. From the recorded intensity, we determine the structure factor $S(\textbf{q})$ by subtracting the solvent background and dividing by the particle form factor determined from dilute suspensions. An example of the angle-averaged structure factor is shown in Fig.~\ref{fig:Sq}, inset. We focus on the first peak of the structure factor to elucidate changes in the nearest-neighbor structure upon yielding.

The applied shear introduces structural anisotropy: in an elastic material, shear leads to a well-known distortion of the structure resulting in an anisotropic intensity distribution along the first ring; this is demonstrated by the angle-resolved structure factor at small strain amplitude in Fig.~\ref{fig:Sq} (red curve). The two-fold (p-wave) symmetry indicates a solid-like response; it reflects the elastic response of the material to local shear distortions~\cite{Argon_79}, consistent with our direct imaging of the strain field by confocal microscopy~\cite{chikkadi_schall11,schall2007}. The latter reveals the ubiquity of quadrupolar elastic fields known as Eshelby field~\cite{Eshelby} associated with elementary shear transformations~\cite{picard2004,schall2007}. The normal strain component of this Eshelby field has a $p$-wave symmetry in the shear plane, which is precisely the symmetry that we observe here. We elucidate the strain dependence of this anisotropy by following the structure factor as a function of increasing strain amplitude. While two-fold symmetry persists at small strain, at larger strain, this symmetry is lost and the structure factor becomes isotropic, as shown by the blue line for $\gamma_0 = 10^0$ in Fig. ~\ref{fig:Sq}.

\begin{figure}
\centering
{\includegraphics[width=0.85\columnwidth]{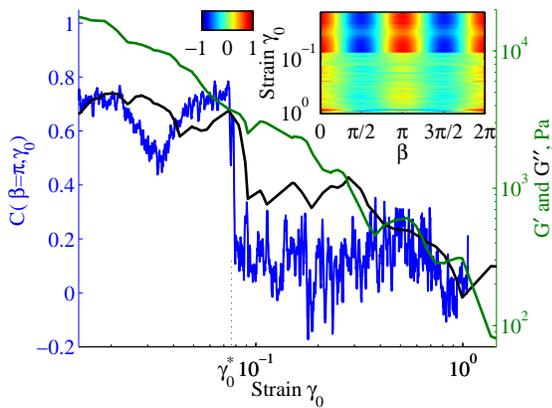}
\caption{(Color online) Order parameter $C(\beta=\pi,\gamma_0)$ as a function of strain amplitude (left axis, blue). Also indicated are the elastic and viscous moduli, $G'$ and $G''$ (right axis, green and black). Inset: Color map showing the evolution of the angular correlation function $C(\beta,\gamma_0)$ with horizontal axis: correlation angle $\beta$; vertical axis: applied strain amplitude $\gamma_0$. Color indicates the value of $C(\beta,\gamma_0)$, see color bar.}
\label{fig:Cor2D_vf30}}
\end{figure}

To highlight this symmetry change most clearly, we determine the angular correlation function of the angle dependent structure factor $S_1(\alpha)$,
\begin{multline}
C(\beta)= \\
\frac{\int_0^{2\pi}(S_1(\alpha+\beta)-<S_1(\alpha)>)(S_1(\alpha)-<S_1(\alpha)>)d\alpha}
{\int_0^{2\pi}(S_1(\alpha)-<S_1(\alpha)>)^2 d\alpha},
\label{eq1}
\end{multline}
where we integrate over all angles $\alpha$ as a function of the correlation angle $\beta$. We reduce possible effects of elliptical distortion of the first ring by averaging radially over an extended range of wave vectors ($\Delta q \sim2w_1$) around $q_1$. This allows us to follow the underlying symmetry most clearly. We illustrate its strain evolution in Fig.~\ref{fig:Cor2D_vf30} inset, where we represent the angular correlation function with color and follow its evolution along the vertical axis. A sudden loss of symmetry is observed at $\gamma_0^{\star}\sim 0.077$, as demonstrated by the sudden disappearance of the p-wave pattern.

To investigate the sharpness of this transition, we define an order parameter that measures the degree of anisotropy. A good choice of such order parameter is the peak value of the correlation function, $C(\beta=\pi)$, which is 1 for the ideal case of p-wave symmetry, and 0 for a complete loss of symmetry. We show this order parameter as a function of $\gamma_0$ in Fig.~\ref{fig:Cor2D_vf30} (blue line). At $\gamma_0^{\star}$, an abrupt drop from $C(\beta=\pi,\gamma_0) \sim 0.8$ to $C(\beta=\pi,\gamma_0) \sim 0$ occurs, demonstrating a surprisingly sharp loss of orientational order and thus melting in the orientational degrees of freedom; at the same time, the mean absolute value of $S(q)$ does not change, indicating robust translational degrees of freedom. A similar symmetry change was observed by us in the real-space imaging of a sheared colloidal glass~\cite{chikkadi_schall11}: the microscopic strain correlations changed symmetry from anisotropic solid to isotropic liquid-like. Such symmetry change reminds of first order equilibrium transitions that demarcate qualitative changes of a material characterized by an order parameter. The order parameter defined here together with the excellent time resolution allow us to indeed demonstrate the sharpness of this transition.

\begin{figure}
\centering
{\includegraphics[width=0.85\columnwidth]{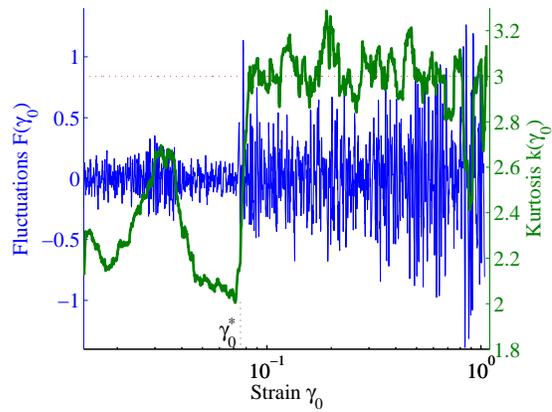}
\caption{(Color online) Time correlation of the order parameter, $F(\gamma_0)$ (left axis, blue curve) and kurtosis $\kappa(\gamma_0)$ (right axis, green curve) as a function of strain amplitude. Red dotted line indicates Gaussian value 3 of the kurtosis. At $\gamma_0^*$, fluctuations increase, and the kurtosis changes sharply to its Gaussian value 3.}
\label{fig:CP2fluct_vf30}}
\end{figure}

Concomitantly with the loss of symmetry, the amplitude of fluctuations increases. We investigate these fluctuations by calculating their time correlation via
\begin{multline}
F(\Delta t)= \\
\frac{1}{T}\int_0^{T}(C(t+\Delta t)-<C(t)>)(C(t)-<C(t)>)dt,
\label{eq2}
\end{multline}
where $t\sim \log(\gamma_0/\gamma_{0min})$ and we correlate order parameter values $C(t)=C(\beta=\pi,t)$ as a function of delay time $\Delta t \sim \Delta\log(\gamma_0)$. Here, $T$ is the averaging time interval. For sufficiently large $T$, the time correlation should pick out the typical time scale of fluctuations, such as for example the underlying oscillation period, during which the colloidal glass may yield and reform \cite{Rogers2011b}. However, our data does not show such characteristic time scale, possibly due to limited resolution. This is confirmed independently by Fourier analysis. We thus interpret these fluctuations as noise. To investigate their amplitude as a function of $\gamma_0$, we choose a short averaging period $T = 1$s; this allows us to clearly observe a sudden increase in the noise amplitude at $\gamma_0^{\star}$ as shown in Fig~\ref{fig:CP2fluct_vf30}.

We elucidate this sudden increase of amplitude by determining the kurtosis $\kappa$. The kurtosis uses higher moments of intensity fluctuations to investigate the Gaussian nature of fluctuations: a kurtosis value of 3 indicates a Gaussian distribution and thus that fluctuations are uncorrelated. We determine instantaneous values of $\kappa$ from spatial fluctuations of the structure factor along the diffraction ring using $\kappa(\gamma_0)=m_4/m_2^2$, where the $i$-th moment of the structure factor $m_i=\frac{1}{2\pi}\int_{0}^{2\pi}{(S_1(\alpha)-<S_1(\alpha)>)^i d\alpha}$. This allows us to follow the kurtosis as a function of applied strain. The resulting evolution of $\kappa$ is shown in Fig.~\ref{fig:CP2fluct_vf30} (green curve). A sharp increase to a value of 3 occurs precisely at $\gamma_0^{\star}$, indicating a sudden transition to Gaussian fluctuations. The evolution of $\kappa$ mirrors the evolution of the order parameter shown in Fig.~\ref{fig:Cor2D_vf30}; hence, the loss of symmetry is accompanied by instantaneous changes to Gaussian intensity distributions. The disappearance of anisotropy, the increase of the amplitude of fluctuations and their transition to Gaussian distributions suggest sudden melting of the glass in the orientational degrees of freedom. We note that the value of strain $\gamma_0^{\star}$ is in reasonable agreement with the value 0.08 reported for the "shear melting" of colloidal glasses~\cite{Weitz2010}; the precise yield strain value, however, may depend weakly on the shear rate and the interactions of the particles.

\begin{figure}
\centering
\subfigure
{\includegraphics[width=0.85\columnwidth]{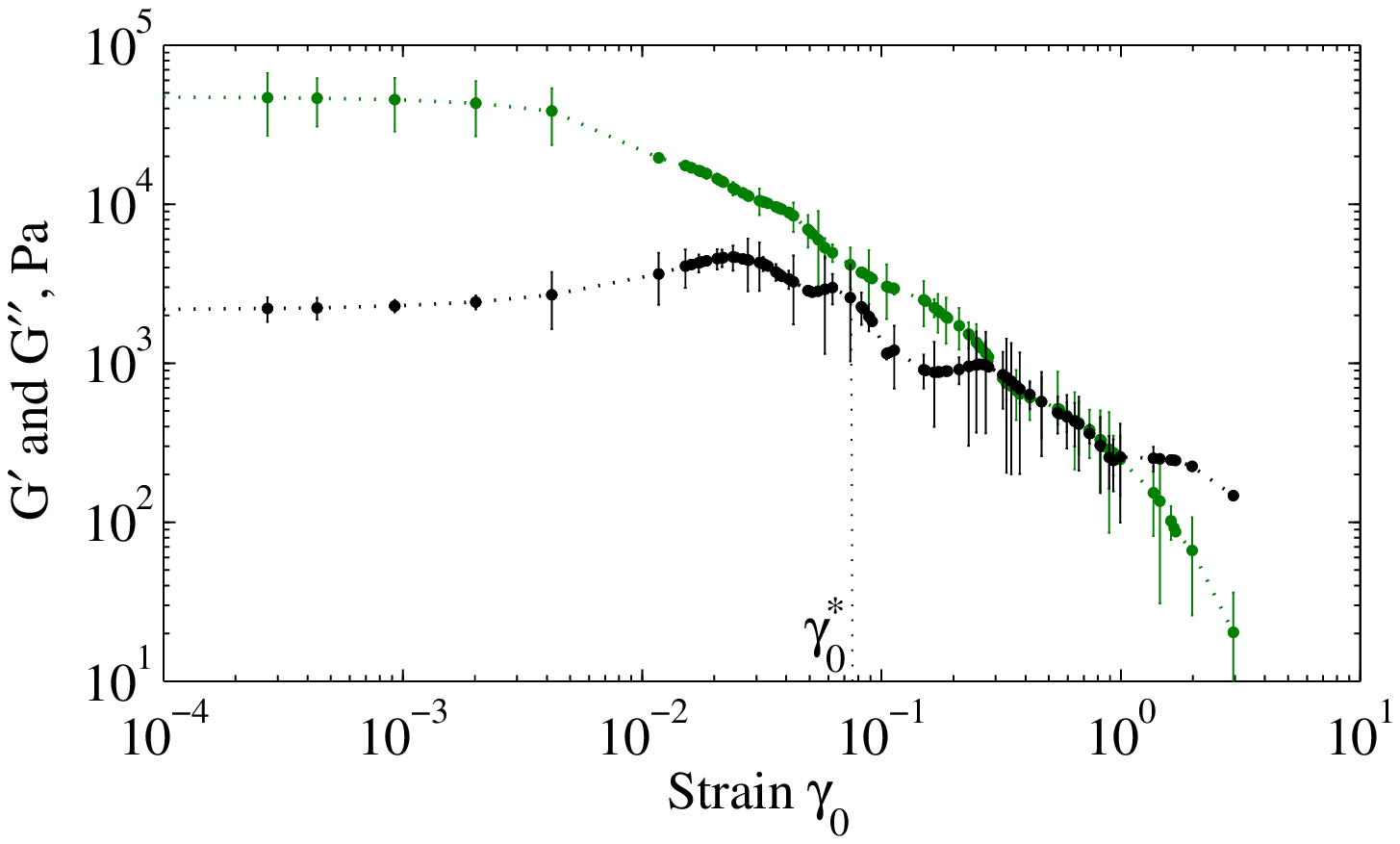}
\label{fig:Gprimes}}
\begin{picture}(0,0)(0,0)
\put(-225,10){(a)}
\end{picture}
\subfigure
{\includegraphics[width=0.85\columnwidth]{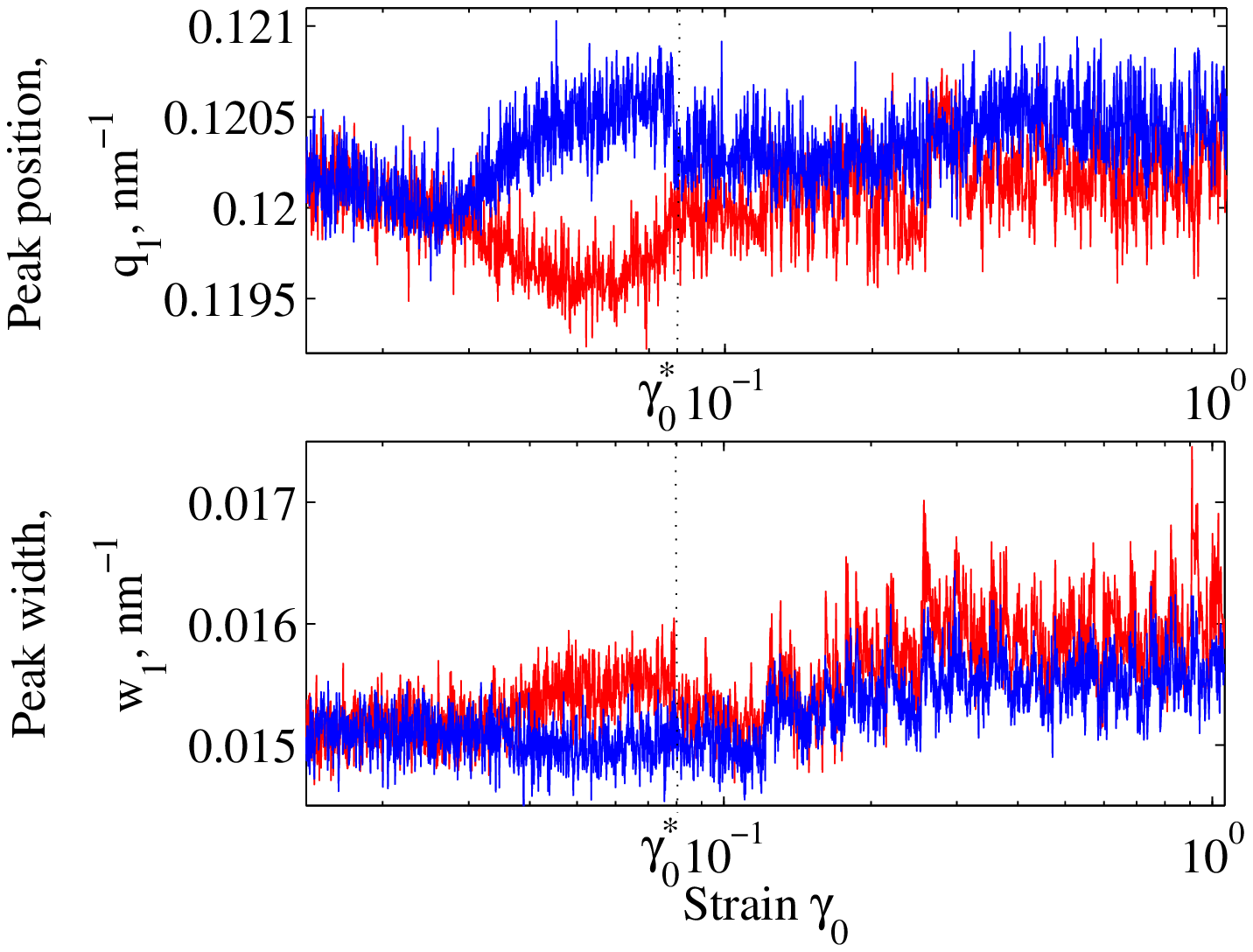}
\label{fig:q1w1_vf30}}
\begin{picture}(0,0)(0,0)
\put(-225,10){(b)}
\end{picture}
\caption{(Color online)
\subref{fig:Gprimes} Elastic and viscous moduli, $G^\prime$ and $G^{\prime\prime}$ as a function of oscillatory strain amplitude.
\subref{fig:q1w1_vf30} Peak position $q_1$ (top) and half-width $w_1$ (bottom) of the first peak of the structure factor as a function of strain amplitude. Red and blue curves correspond to directions along and perpendicular to the shear, respectively.}
\end{figure}

We thus identify a sharp, dynamically induced transition in the glass structure. To link this microscopic transition to the rheological behavior of the glass, we follow the strain-dependant storage and loss moduli, $G^\prime$ and $G^{\prime\prime}$, simultaneously with the structure factor (see Fig.~\ref{fig:Gprimes}). These moduli show the well-known strain dependence of dense suspensions: the plateau at small strain with $G^\prime > G^{\prime\prime}$ is followed by the decrease of the moduli and their intersection, indicating the non-linear regime. The point where the two moduli cross is generally associated with the yield point~\cite{BonnDenn}: the storage modulus decreases below the loss modulus indicating a gradual loss of elasticity.
Interestingly, the point where the two curves meet is close to $\gamma_0^{\star}$, allowing us to associate the sharp structural transition with the rheological yielding of the material. This is shown most clearly in Fig.~\ref{fig:Cor2D_vf30}, where the moduli have been reproduced in enlarged form (green and black lines); the exact intersection of $G^\prime$ and $G^{\prime\prime}$, however, is hard to pinpoint and the curves even coexists for some range around $\gamma_0\sim10^{-1}$. Surprisingly, our structure factor analysis provides us with a much sharper definition of the yielding point ($\gamma_0^{\star}$), which makes the structural correlation analysis a powerful tool to pinpoint the yielding of the material. We note, however, that the moduli $G^\prime$ and $G^{\prime\prime}$ as usual represent only the first harmonic response; higher harmonics are missing in this representation, although they can be quite significant~\cite{Wilhelm,Laettinga}, and their inclusion might provide a sharper mechanical signature of yielding.

Further signature of yielding is observed in the position $q_1$ and width $w_1$ of the first peak as shown in Fig.~\ref{fig:q1w1_vf30}. The initial solidity of the material reflects in the anisotropy of particle distances; the decrease of $q_1$ along (red curve), and its increase perpendicular to the shear direction (blue curve) indicates that particle separations increase along and decrease perpendicular to the applied shear, making them move past each more easily. The concomitant increase in the peak width (Fig.~~\ref{fig:q1w1_vf30}, bottom) indicates that this is accompanied by slight shear-induced disordering. This anisotropy increases until $\gamma_0\sim\gamma_0^{\star}$, where it suddenly disappears: the material can no longer sustain the anisotropic structure, and changes spontaneously into an isotropic fluid-like state. We note that the anisotropy shown in Fig.~\ref{fig:q1w1_vf30} appears as dip in the correlation function (Fig.~\ref{fig:Cor2D_vf30}) and as peak in the kurtosis (Fig.~\ref{fig:CP2fluct_vf30}); this allows us to estimate the effect of ring distortions - both real and artificial - on the presented analysis. We conclude that while ring distortions have a visible effect, the observed sharp transition at $\gamma_0^{\star}$ cannot be explained by these continuously evolving distortions and indicates a real structural transition of the glass. This abrupt transition, characterized by the order parameter $C(\beta=\pi)$, reminds of first-order transitions accompanying conventional solid-liquid phase changes. In the case presented here the transition is dynamically induced, triggered by the application of an external stress field to the glass.

By combining oscillatory rheology and time-resolved x-ray scattering, we have identified a sharp structural transition at the yielding of a glass. The structural anisotropy characteristic of a solid vanishes abruptly, and isotropic Gaussian fluctuations characteristic of a liquid appear, indicating a sharp dynamically induced transition from a solid to a liquid-like state. While the overall structural effect is small and difficult to detect by real-space techniques such as confocal microscopy, the large averaging power of x-ray scattering allows us to identify this transition clearly. The definition of angular correlation functions as order parameters allowed us to pinpoint the transition and demonstrate its sharpness. This transition, induced by application of an external shear field, thus looks akin to conventional first-order transitions.

\section{Acknowledgements}

The authors thank T. A. Nguyen for her support during the measurements and G. Petekidis and J. Dhont for useful discussions. We thank DESY, Petra III, for access to the x-ray beam. This work was supported by the Foundation for Fundamental Research on Matter (FOM) which is subsidized by the Netherlands Organisation for Scientific Research (NWO). P.S. acknowledges support by a Vidi fellowship from NWO.

\end{document}